\documentclass[epj]{webofc}
\usepackage[varg]{txfonts}   % Web of Conferences font
\usepackage[alsoload=hep]{siunitx}
\usepackage{tikz}

\DeclareSIUnit\gev{\giga\eV}
\DeclareSIUnit\gevperc{\giga\eV\per\clight}
\DeclareSIUnit\gevpercsq{\giga\eV\per\clight\squared}

\woctitle{MESON2018 - the 15$^\textrm{th}$ International Workshop on Meson Physics}

\newcommand{\pnn}{K\to\pi\nu\bar{\nu}}
\newcommand{\pnnc}{K^+\to\pi^+\nu\bar{\nu}}
\newcommand{\pnnn}{K^0\to\pi^0\nu\bar{\nu}}
\newcommand{\group}[1]{\left(#1\right)}
\newcommand{\sqgroup}[1]{\left[#1\right]}
\newcommand{\abs}[1]{\left|#1\right|}

\begin{document}
\selectlanguage{english}
\title{Results and prospects for $\pnn$ at NA62 and KOTO}

% insert email only for speaker/presenter
\author{Nicolas~Lurkin\inst{1}\fnsep\thanks{\email{nicolas.lurkin@cern.ch}} 
% comment out the next line if not needed
       \\for the NA62 Collaboration%\thanks{R. Aliberti, F. Ambrosino, R. Ammendola, B. Angelucci, A. Antonelli, G. Anzivino, R. Arcidiacono, M. Barbanera, A. Biagioni, L. Bician, C. Biino, A. Bizzeti, T. Blazek, B. Bloch-Devaux, V. Bonaiuto, M. Boretto, M. Bragadireanu, D. Britton, F. Brizioli, M.B. Brunetti, D. Bry-man, F. Bucci, T. Capussela, A. Ceccucci, P. Cenci, V. Cerny, C. Cerri, B. Checcucci, A. Conovaloff, P. Cooper, E. Cortina Gil, M. Corvino, F. Costantini, A. Cotta Ramusino, D. Coward, G. D’Agostini, J. Dainton, P. Dalpiaz, H. Danielsson, N. De Simone, D. Di Filippo, L. Di Lella, N. Doble, B. Dobrich, F. Duval, V. Duk, J. Engelfried, T. Enik, N. Estrada-Tristan, V. Falaleev, R. Fantechi, V. Fascianelli, L. Federici, S. Fedotov, A. Filippi, M. Fiorini, J. Fry, J. Fu, A. Fucci, L. Fulton, E. Gamberini, L. Gatignon, G. Georgiev, S. Ghinescu, A. Gianoli, M. Giorgi, S. Giudici, F. Gonnella, E. Goudzovski, C. Graham, R. Guida, E. Gushchin, F. Hahn, H. Heath, T. Husek, O. Hutanu, D. Hutchcroft, L. Iacobuzio, E. Iacopini, E. Imbergamo, B. Jenninger, K. Kampf, V. Kekelidze, S. Kholodenko, G. Khoriauli, A. Khotyantsev, A. Kleimenova, A. Korotkova, M. Koval, V. Kozhuharov, Z. Kucerova, Y. Kudenko, J. Kunze, V. Kurochka, V. Kurshetsov, G. Lanfranchi, G. Lamanna, G. Latino, P. Laycock, C. Lazzeroni, M. Lenti, G. Lehmann Miotto, E. Leonardi, P. Lichard, L. Litov, R. Lollini, D. Lomidze, A. Lonardo, P. Lubrano, M. Lupi, N. Lurkin, D. Madigozhin, I. Mannelli, G. Mannocchi, A. Mapelli, F. Marchetto, R. Marchevski, S. Martellotti, P. Massarotti, K. Massri, E. Maurice, M. Medvedeva, A. Mefodev, E. Menichetti, E. Migliore, E. Minucci, M. Mirra, M. Misheva, N. Molokanova, M. Moulson, S. Movchan, M. Napolitano, I. Neri, F. Newson, A. Norton, M. Noy, T. Numao, V. Obraztsov, A. Ostankov, S. Padolski, R. Page, V. Palladino, C. Parkinson, E. Pedreschi, M. Pepe, M. Perrin-Terrin, L. Peruzzo, P. Petrov, F. Petrucci, R. Piandani, M. Piccini, J. Pinzino, I. Polenkevich, L. Pontisso, Yu. Potrebenikov, D. Protopopescu, M. Raggi, A. Romano, P. Rubin, G. Ruggiero, V. Ryjov, A. Salamon, C. Santoni, G. Saracino, F. Sargeni, V. Semenov, A. Sergi, A. Shaikhiev, S. Shkarovskiy, D. Soldi, V. Sougonyaev, M. Sozzi, T. Spadaro, F. Spinella, A. Sturgess, J. Swallow, S. Trilov, P. Valente, B. Velghe, S. Venditti, P. Vicini, R. Volpe, M. Vormstein, H. Wahl, R. Wanke, B. Wrona, O. Yushchenko, M. Zamkovsky, A. Zinchenko.}
}

\institute{University of Birmingham, Birmingham, United Kingdom}

\abstract{The $\pnn$ ultra-rare decays are precisely computed in the Standard Model (SM) and are ideal probes for physics beyond the SM. The NA62 experiment at the CERN SPS is designed to measure the charged channel with a precision of \num{10}{\%}. The statistics collected in 2016 allows to reach the SM sensitivity. The KOTO experiment at J-PARC aims at reaching the SM sensitivity before performing a measurement with $\sim100$ signal events. The NA62 preliminary result for the charged channel is presented, together with the current experimental status of the neutral channel and their prospects for the coming years.
}
\maketitle
\section{Introduction}
\label{intro}
Due to its strong suppression, the $\pnn$ decay is a golden channel for precision tests of the Standard Model (SM) and search for physics Beyond the Standard Model (BSM). This decay is a flavour changing neutral current, forbidden at tree level, proceeding through box and electromagnetic penguin diagrams. It also benefits from an additional suppression from the CKM matrix element and a quadratic GIM mechanism. 

This ultra-rare Kaon decay is also theoretically very clean. It is described mostly by a short-distance effective Hamiltonian receiving contributions from the top quark loop, with small contribution from the charm quark loop and long-distance corrections. The hadronic matrix element can be extracted from the well measured, isospin rotated $K^+\to\pi^0 e^+\nu$ decay. Overall the uncertainties on the CKM matrix elements are dominating the theoretical error budget~\cite{Buras2015}. The remaining relative uncertainties are about \SI{3.6}{\percent} (\SI{1.5}{\percent}) for the charged (neutral) channel, and the continuous improvement on the precision of the CKM parameters therefore enables to further reduce the total uncertainties on the branching ratios. The latest numerical evaluation leads to:
\begin{alignat}{2}
\mathcal{B}(\pnnc) &= \num{8.39(30)e-11}\cdot \sqgroup{\frac{\abs{V_{cb}}}{\num{40.7e-3}}}^{2.8}\sqgroup{\frac{\gamma}{\ang{73.2}}}^{0.74} \\
&= \num{8.4(10)e-11} \\
\mathcal{B}(\pnnn) &= \num{3.36(5)e-11} 
\cdot \sqgroup{\frac{\abs{V_{ub}}}{\num{3.88e-3}}}^{2}\sqgroup{\frac{\abs{V_{cb}}}{\num{40.7e-3}}}^2\sqgroup{\frac{\sin{\gamma}}{\sin{\ang{73.2}}}}^{0.74}\\
&= \num{3.4(6)e-11}
\end{alignat}
%\begin{alignat}{2}
%\mathcal{B}(\pnnc) &= \num{8.4(10)e-11}\\
%\mathcal{B}(\pnnn) &= \num{3.4(6)e-11}
%\end{alignat}

In this context even small BSM effects could have a significant impact on the branching ratio. It is also interesting to exploit the correlations between the charged and the neutral channel as they vary significantly between different classes of models introducing BSM physics (Custodial Randall-Sundrum~\cite{Blanke2009}, MSSM analysis~\cite{Isidori2006, Salam2014}, simplified $Z$,$Z'$ models~\cite{Buras2015_2}, littlest Higgs with T-parity~\cite{Blanke2016}, lepton flavour violation models~\cite{Bordone2017}). The combination of the measurements for $\pnnc$ and $\pnnn$, but also with the their $B$ physics counterparts ($B\to K\nu\bar{\nu}$), can lead to strong constraints on these models.

The experimental situation is very different from the theoretical one. For $\pnnc$, the only measurement is extracted from seven event candidates at the E787 and E949 experiments at BNL~\cite{bnl2008,bnl2009}. For $\pnnn$ only an upper limit is available, where the strongest comes from the E391 experiment at KEK~\cite{kek2010}. The measurements are:
\begin{alignat}{2}
\mathcal{B}(\pnnc) &= \group{17.3^{+11.5}_{-10.5}}\times 10^{-11}\\
\mathcal{B}(\pnnn) &< \num{2.6e-8}\ \text{(90\% C.L.)}
\end{alignat}
The relative uncertainty on the charged measurement are of the order of \SI{60}{\percent}, which does not allow to conclude on a possible discrepancy with the predictions. The neutral measurement is still an order of magnitude higher than the Grossman-Nir bounds~\cite{grossman1997}, which limits the ratio of $\mathcal{B}(\pnnn)/\mathcal{B}(\pnnc)$  based on CP violation considerations.

\section{The NA62 experiment}
\label{na62}
The fixed target NA62 experiment at the CERN SPS aims at measuring $\mathcal{B}(\pnnc)$ with a precision of 10\% using an in-flight decay technique. In order to achieve this goal a total of \num{e13} kaon decays should be collected in a few years of data taking, with a maximum of 10\% background contamination in the final signal sample. The primary \SI{400}{\gevperc} proton beam impinges on a Beryllium target, producing secondary particles. The secondary beam optics selects, collimates and focuses particles with a momentum of \SI{75}{\gevperc} to the \SI{60}{\m} long evacuated decay volume. The beam is composed of 70\% pions, 23\% protons and 6\% kaons.

The signal consists of one incoming kaon track and one outgoing pion track, with no other activity in the detector. The main background to the signal comes from the kaon decay channels with highest branching ratio: principally $K^+\to\mu^+\nu(\gamma)~(K_{\mu\nu(\gamma)})$ and $K^+\to\pi^+\pi^0(\gamma)~(K_{\pi\pi(\gamma)})$. There are also contributions from upstream decays in the beam line and interactions between the beam particles and upstream detectors. The high background rejection required is provided by combining various techniques: kinematic suppression, high efficiency veto system, time resolution.

\begin{figure}
	\centering
	\includegraphics[width=0.9\textwidth]{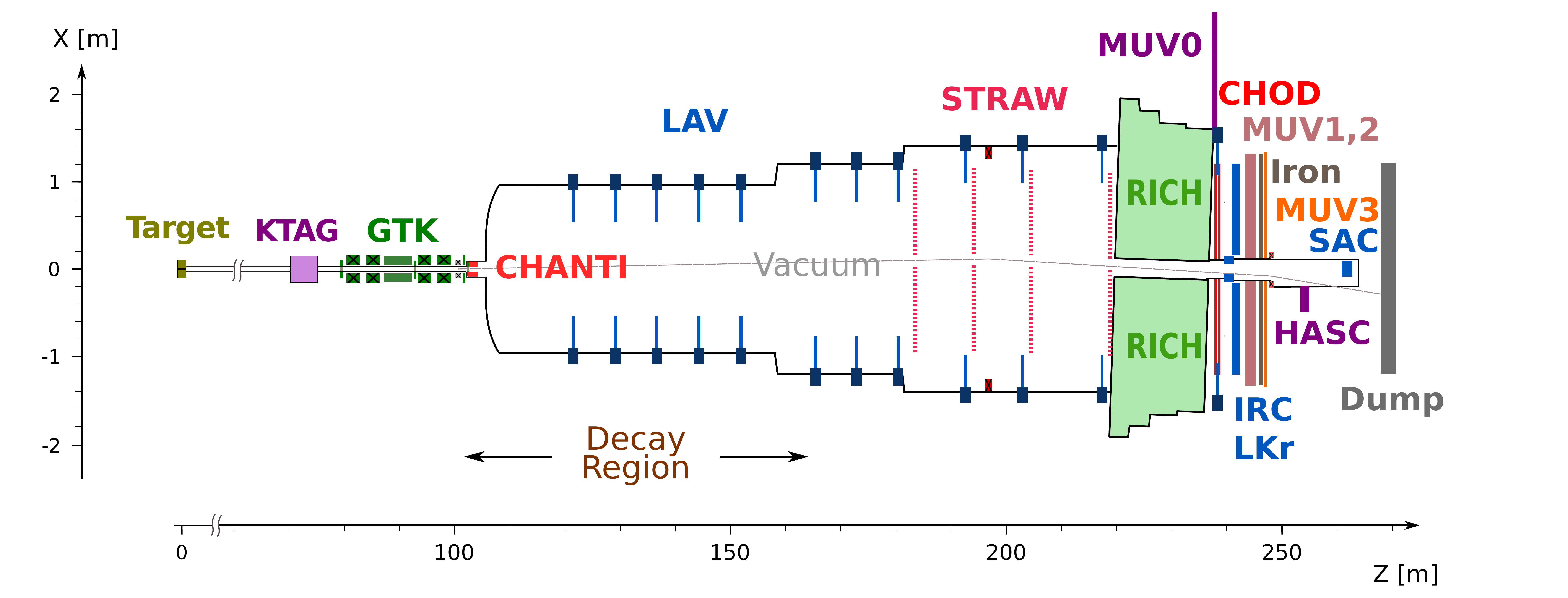}
	\caption{Schematic view of the NA62 experimental set-up}
	\label{fig:detector}
\end{figure}

The full schematic of the detector can be seen in Fig.~\ref{fig:detector}, and a detailed description of the experimental set-up can be found in \cite{cortina2017}. A differential Cherenkov counter (KTAG) identifies kaons in the beam line with a time resolution of ${\sim}\SI{100}{\ps}$, which is matched to downstream activity to reject beam induced background. It is followed by the beam spectrometer (GTK) measuring the kaon momentum. Immediately following the decay volume, a straw spectrometer (STRAW) measures the downstream tracks momentum. A Ring Imaging Cherenkov detector (RICH) provides particle identification for $\pi^\pm,\mu^\pm, e^\pm$ in the momentum range \SIrange{15}{35}{\gevperc}. These measurements allow to build the squared missing mass variable $m_\text{miss}^2 \equiv (p_K - p_\pi)^2$, where $p_K$ and $p_\pi$ are the 4-momenta of the beam and downstream tracks under the kaon and pion hypothesis, respectively. This variable is used to discriminate between the signal and background kinematics, allowing a \num{e4} rejection factor. Finally a set of highly efficient veto for muons, photons and inelastic interaction provide an almost hermetic coverage reaching a \num{e7} suppression factor.

\subsection{The $K^+\to\pi^+\nu\bar{\nu}$ analysis}
The events are reconstructed according to the signal signature described above: a single track topology is selected by spatially matching a reconstructed track in the STRAW with signals in the CHOD and RICH detectors.
%The time resolution on the track is about \SI{100}{\ps}. 
A $K^+$ identified in the KTAG detector and traced in the GTK is matched in time with the upstream track. The kaon decay vertex is built at the intersection of the two tracks and is required to be in a \SI{50}{\m} long region starting \SI{10}{\m} downstream of the last GTK station. The $\pi^+$ identification is performed using in combination the RICH and the calorimeters, providing a total muon rejection of \num{e-8} for a $\pi^+$ efficiency of \SI{64}{\%}, measured on independent samples of kinematically selected $K_{\pi\pi}$ and $K_{\mu\nu}$. The remaining background is principally $K_{\pi\pi}$ decays, which are further suppressed using the electromagnetic calorimeters (LKr, SAC, IRC, LAVs). Events with extra energy deposits in-time with the $\pi^+$ track 
%(ranging between \SI{\pm3}{\ns} and \SI{\pm10}{\ns}) 
are rejected, providing a $\pi^0$ suppression of \num{3e-8}. 
The signal region defined in the $m_\text{miss}^2$ variable is driven by its resolution (\SI{e-3}{\gev\squared\per\clight\tothe{4}}) and is divided into two sub-regions (R1 and R2) below and above the $K_{\pi\pi}$ peak respectively. The 4-momenta used in the squared missing mass computation can be reconstructed using different methods: using STRAW or RICH for $p_\pi$, and using GTK or the nominal beam parameters for $p_K$. Imposing constraints on $m_\text{miss}^2$ reconstructed using various combinations of these methods protects against mis-reconstruction of the momenta.
%A small remaining background contributions from kaon decays are also entering the signal region: $K_{\pi\pi}$ and $K_{\mu\nu}$ through non-Gaussian resolution and radiative tails, $K^+\to\pi^+\pi^+\pi^-~(K_{\pi\pi\pi})$ through non-Gaussian resolution, and $K^+\to\pi^+\pi^- e^+ \nu~(K_{e4})$ and $K^+\to\ell^+ \pi^0\nu~(K_{l3})$ with a neutrino in the final state. 
A sample selected using calorimeters only is used to measure the kinematic suppression factors, which is \num{1e-3} for $K_{\pi\pi}$ and $\num{3e-4}$ for $K_{\mu\nu}$. The expectations for the final background contamination are verified in control regions on the side of the signal regions R1 and R2 for $K_{\pi\pi(\gamma)}$ and $K_{\mu\nu}$, or estimated from MC for 
%$K_{e4}$
$K^+\to\pi^+\pi^- e^+ \nu~(K_{e4})$. The final acceptance for the signal, extracted from Monte-Carlo (MC) simulations, is \SI{1}{\%} (\SI{3}{\%}) in R1 (R2), for a total acceptance of $A_{\pi\nu\bar\nu} = \SI{4}{\%}$.

The Single Event Sensitivity ($SES$) is defined as $SES = 1/\left(N_K\cdot A_{\pi\nu\bar\nu}\cdot\varepsilon_\text{Trig} \cdot\varepsilon_\text{RV}\right)$ where $N_K$, is the number of kaon decays, $\varepsilon_\text{Trig}$ is the $\pi\nu\bar{\nu}$ trigger efficiency and $\varepsilon_\text{RV}$ is the signal efficiency resulting from the rejection of events due to accidental activity in the detector. The number of kaon decays $N_K=\num{1.21(2)e11}$ is measured from a control-triggered sample of $K_{\pi\pi}$ selected using a $\pi\nu\bar{\nu}$-like selection on which the final $\gamma$, multiplicity and $m_\text{miss}^2$ cuts are not applied. 
%The acceptance for this sample is \SI{10}{\%}. 
The trigger efficiency $\varepsilon_\text{Trig}$ is measured to be \SI{88}{\%} using control data. The random veto efficiency $\varepsilon_\text{RV}$ depends on the beam intensity and is evaluated to \num{0.76(4)} from a sample of $K_{\mu\nu}$ candidates. The $SES = \group{3.15\pm 0.01_\text{stat}\pm 0.24_\text{syst}}\times 10^{-10}$ for the data sample analysed is dominated by the systematic uncertainties, which are shown in Table~\ref{tab:syst}.

\begin{table}[!htb]
	\centering
	\caption{Summary of the systematic uncertainties for the $SES$ computation.}
	\label{tab:syst}
	\begin{tabular}{|l|c|}
		\hline
		Source                             & $\delta SES(\num{e-10})$ \\ \hline
		Random Veto                        &        $\pm0.17$         \\
		$N_K$                              &        $\pm0.05$         \\
		Trigger efficiency                 &        $\pm0.04$         \\
		Definition of $\pi^+\pi^0$         &        $\pm0.10$         \\
		Momentum spectrum                  &        $\pm0.01$         \\
		Simulation of $\pi^+$ interactions &        $\pm0.09$         \\
		Extra activity                     &        $\pm0.02$         \\
		GTK Pileup simulation              &        $\pm0.02$         \\ \hline
		Total                              &        $\pm0.24$         \\ \hline
	\end{tabular}
\end{table}

\begin{table}[!htb]
	\centering
	\caption{Summary of the expected number of signal and background events in the signal regions R1 and R2 for the 2016 analysis.}
	\label{tab:expected}
	\begin{tabular}{|l|l|}
		\hline
		Process                        & Expected events (R1+R2)                                                   \\ \hline
		$K^+\to\pi^+\nu\bar{\nu}$ (SM) & $0.267 \pm 0.001_\text{stat} \pm 0.020_\text{syst} \pm 0.032_\text{ext}$  \\
		Total Background               & $0.15  \pm 0.09_\text{stat}  \pm 0.01_\text{syst}$                        \\ \hline
		$K^+\to\pi^+\pi^0 (\gamma)$ IB & $0.064 \pm 0.007_\text{stat} \pm 0.006_\text{syst}$                       \\
		$K^+\to\mu^+\nu   (\gamma)$ IB & $0.020 \pm 0.003_\text{stat} \pm 0.003_\text{syst}$                       \\
		$K^+\to\pi^+\pi^- e^+ \nu$     & $0.018\left. ^{+0.024}_{-0.017}\right|_\text{stat} \pm 0.009_\text{syst}$ \\
		$K^+\to\pi^+\pi^+ \pi^-$       & $0.002 \pm 0.001_\text{stat} \pm 0.002_\text{syst}$                       \\
		Upstream background            & $0.050\left. ^{+0.090}_{-0.030}\right|_\text{stat}$                       \\ \hline
	\end{tabular}
\end{table}

\begin{figure}
	\centering
 	\begin{tikzpicture}
		\draw (0,0) node{\includegraphics[width=.7\textwidth]{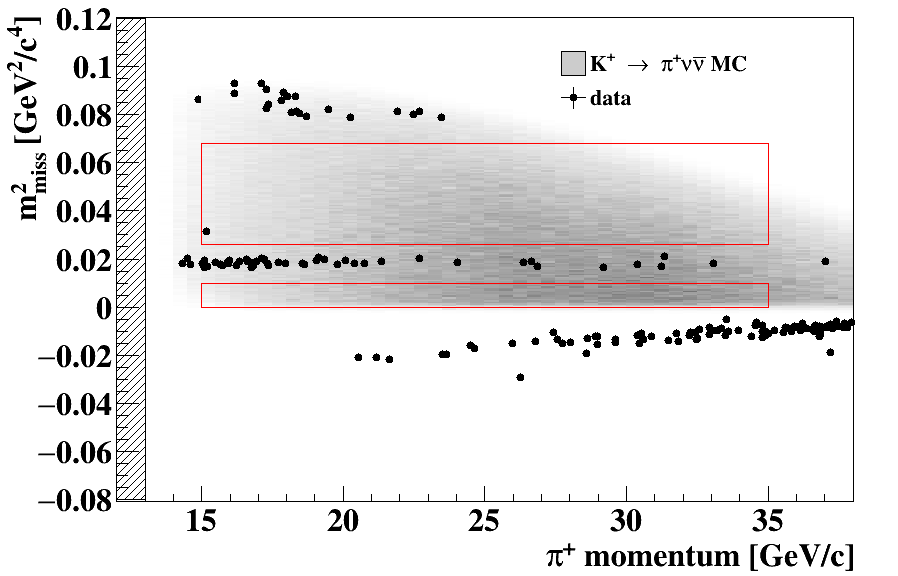}};
		\draw[red,thick] (-2.2,1.2) node {R1};
		\draw[red,thick] (-2.2,-0.1) node {R2};
	\end{tikzpicture}
	\caption{
		Distribution of the $m^2_\text{miss}$ variable as a function of $p_{\pi^+}$. The dots are data events passing the $\pi\nu\bar{\nu}$ selection, except the cuts on $m_\text{miss}^2$ and $p_{\pi^+}$. The grey area corresponds to the density of MC events. The red lines define the signal regions R1 and R2. One event is observed in R2.
	}
	\label{fig:na62_result}
\end{figure}

\subsection{NA62 result and prospect}
After the full selection, one event is found in R2 after un-blinding the signal region. The summary of the expected signal and background events in the signal regions is shown in Table~\ref{tab:expected}. From this, an upper limit on the branching fraction of $\pnnc$ can be set using the $\text{CL}_s$ method: $
%\begin{alignat}{2}
	\mathcal{B}(\pnnc) < \num{10e-10}\ \text{(90\% C.L.)}%\\
%	\mathcal{B}(\pnnc) &< \num{14e-10}\ \text{(95\% C.L.)}
%\end{alignat}
$.
The result presented corresponds to the analysis of data taken in 2016, while approximately 20 times more data have been collected in 2017. Improvements on the signal acceptance, background reduction and reconstruction efficiency are also expected. With another year of data taking in conditions similar to 2017, a total of about 20 SM $\pnnc$ events is expected.

\section{The KOTO experiment}
The KOTO experiment at J-PARC is complementary to NA62, both in its goal and in the technique used. The first phase of the experiment aims at observing a $\pnnn$ signal, before increasing the sensitivity and moving to the second phase with a measurement of $\mathcal{B}(\pnnn)$ from about 100 events. 
A secondary $K_L$ beam with \SI{1.4}{\gevperc} peak energy is used by the experiment.
%A primary \SI{30}{\gevperc} beam is impinging on a gold target and the secondary low energy $K_L$ beam, with \SI{1.4}{\gevperc} peak energy, extracted at \SI{16}{\degree} is used by the experiment.
The neutral kaons are allowed to decay in a \SI{3}{\m} long fiducial region evacuated at \SI{5e-7}{\bar}.

The experimental signature consists only in detecting two photons with missing transverse momentum. The photon energy is measured by a Cesium Iodide calorimeter (CsI) located downstream of the decay volume. By assuming the two $\gamma$ come from a neutral pion, the longitudinal vertex position $Z_\text{vtx}$ and the $\pi^0$ missing transverse momentum $P_t$ can be computed and define the signal region. 
%The signal region is carefully selected in the plane defined by these two variables to reject background. 
The fiducial volume is further surrounded by veto systems to reject events with additional particles.

\subsection{KOTO results and prospects}
The collaboration published results from the analysis of about 100 hours of run taken in 2013~\cite{ahn2017}. The number of kaons collected is $N_K = \num{2.4e11}$, which corresponds to $SES=\num{1.3e-8}$. The upper limit set on the branching fraction is $
%\begin{equation}
	\mathcal{B}(\pnnn) < \num{5.1e-8}\ (\num{90}{\%}~\text{C.L.})\ .
%\end{equation}
$
The main background sources are $K_L\to\pi^+\pi^-\pi^0$ decays and halo neutrons interactions with the detector. Dedicated runs with an aluminium target have been performed to improve the discrimination between neutron and photon clusters in the CsI calorimeter. This resulted in a factor 5 improvement on the reduction of neutron induced background. Other improvements on the experimental system 
%(better beam alignment, beam profile monitor, thinner vacuum window) 
further reduce the impact of neutron interactions. Finally the installation of a new beam pipe charged veto allowed to reduce the acceptance loss by \SI{40}{\%}.

A preliminary result with \SI{60}{\%} more statistics and the improvements mentioned above indicate a $SES\sim\num{5.9e-9}$ with reduced background. The analysis of the full 2015-2016 dataset should bring the $SES$ below \num{e-9}. Further upgrades of the detector and beam line are planned to bring the sensitivity down to the SM level.

\section{Conclusions}
The $\pnn$ ultra-rare decays are excellent probes for new physics. Their branching ratios are both currently known to a very good precision in the SM, with uncertainties mostly arising from the precision on the CKM matrix elements. The analysis performed at the NA62 experiment reports one observed event with \num{0.27} expected SM signal event and \num{0.15} background events. This result validates the chosen decay-in-flight technique and allows to set a limit at \SI{90}{\%} C.L. of $\mathcal{B}(\pnnc) < \num{10e-10}$. The KOTO experiment published a result which sets the limit $\mathcal{B}(\pnnn) < \num{5.1e-8}$ at \SI{90}{\%} C.L. for the neutral channel. Both collaborations are continuing to improve their experimental set-up and to take data. 
%The experimental situation should greatly improve in the coming years.

%
% BibTeX or Biber users please use (the style is already called in the class, ensure that the "woc.bst" style is in your local directory)
 \bibliography{bibli.bib}
%\begin{thebibliography}{00}
%F.~Author~\textit{et al.}, Journal \textbf{Volume}, page numbers (year) {\em{no trailing dot!}}
%B.~Author, \textit{Book title} (Publisher, place, year) page numbers {\em{no trailing dot!}}
% etc
%\end{thebibliography}

\end{document}